\documentclass[preprint,twocolumn,10pt,a4paper,byrevtex,prb,unsortedaddress,superscriptaddress,longbibliography]{revtex4-1}

\usepackage{amssymb}
\usepackage{natbib}
\usepackage{graphicx}
\usepackage{amsmath}
\usepackage{float}
\usepackage{verbatim}
\usepackage{color}
\usepackage{bbold} 
\usepackage{hyperref}
\begin{document}

\definecolor{Red}{rgb}{1,0,0}
\definecolor{Blue}{rgb}{0,0,0}
\definecolor{Red_2}{rgb}{1,0,0}
\newcommand{\farkhad}[1]{\textcolor{Blue}{#1}}
\newcommand{\konstantin}[1]{\textcolor{Blue}{#1}}
\definecolor{Orange}{rgb}{1,0.55,0}
\newcommand{\diego}[1]{\textcolor{Blue}{#1}}
\newcommand{\diegotwo}[1]{\textcolor{Blue}{#1}}

\title{Dynamics and reversible control of the vortex Bloch-point vortex domain wall in short cylindrical magnetic nanowires}

\author{Diego Caso}

\author{Pablo Tuero}%

\author{Javier García}
\affiliation{ 
Departamento F\'isica de la Materia Condensada C03, Universidad Aut\'onoma de Madrid, Madrid 28049, Spain 
}%

\author{Konstantin Y. Guslienko}
\affiliation{%
Departamento Pol\'imeros y Materiales Avanzados: F\'isica, Qu\'imica y Tecnolog\'ia, Universidad del Pa\'is Vasco, UPV/EHU, 20018 San Sebasti\'an, Spain
}%
\affiliation{EHU Quantum Center, University of the Basque Country, UPV/EHU, 48940 Leioa, Spain}
\affiliation{IKERBASQUE, the Basque Foundation for Science, 48009 Bilbao, Spain}
\date{\today}
\author{Farkhad G. Aliev}%
\email{farkhad.aliev@uam.es}
\affiliation{ 
Departamento F\'isica de la Materia Condensada C03, Universidad Aut\'onoma de Madrid, Madrid 28049, Spain 
}%
\affiliation{IFIMAC and INC, Universidad Aut\'onoma de Madrid, Madrid 28049, Spain}
\begin{abstract}
 \diego{Fast and efficient} 
 switching of nanomagnets is one of the main challenges in the development of future magnetic memories. We numerically investigate the evolution of the static and dynamic spin wave (SW) magnetization in short (50-400 nm length and 120 nm diameter) cylindrical ferromagnetic nanowires, where competing single vortex (SV) and vortex domain wall with a Bloch point (BP-DW) magnetization configurations could be formed. For a limited nanowire length range (between 150 and 300 nm) we demonstrate a reversible microwave field induced (forward) and opposite spin currents (backwards) transitions between the topologically different SV and BP-DW states. By tuning the nanowire length, excitation frequency, the microwave pulse duration and the spin current values we show that the optimum  (low power) manipulation of the BP-DW could be reached by a microwave excitation tuned to the main SW mode and for nanowire lengths around 230-250 nm, where single vortex domain wall magnetization reversal via nucleation and propagation of SV-DW takes place. An analytical model for dynamics of the Bloch point provides an estimation of the gyrotropic mode frequency close the one obtained via micromagnetic simulations. \diego{A practical implementation of the method on a device has been proposed involving microwave excitation and the generation of the opposite spin currents via spin orbit torque.} Our findings open a new pathway for the creation of unforeseen topological magnetic memories.
\end{abstract}

\maketitle

\section{\label{sec:level1}Introduction}

In the last decades, numerous advances have been made in the development of spin based data storage and information processing devices. Low dimensional structures with a confined magnetic vortex have been in focus as the promising \farkhad{candidate} in the creation of novel types of quantum magnetic memories.  Their functionality as magnetic memories is based on the control of the change in their topology, mainly between two distinct magnetization configurations. For example, the control of the polarity of magnetic vortices created in nanometric disks (i.e. with functionality primarily restricted to the material plane \cite{Kammerer2011}) through the excitation of spin waves (SW) \cite{Aliev2009, Awad2010,Aliev2011}  has been demonstrated. \farkhad{A great advantage of using the \diego{excitation} of SWs compared with spin transfer torque (STT)} \cite{Chureemart2011}, spin orbit torque (SOT) \cite{Kohno2020} or magnetic field \cite{Sun2022}
 \farkhad{for the topology control is the possibility of fine control of the topology of each individual element by tuning the excitation frequency.}
\farkhad{So far, the most widely used method to drive locally the DWs has been by injection of the spin currents \cite{Wong2016, Weiser2010}. }
 
\farkhad{Emergent three-dimensional (3D) magnetic structures \cite{Raftrey2022}} allow a greater variety of textures and magnetic topologies to be achieved, leading to \farkhad {new} effects \cite{Pacheco2017, Sobucki2022}. An example of this is the creation of tunnels of chiral "skyrmions" and "bobbers", more complex 3D magnetic textures in nanoribbons \cite {Zheng2018}. Topological singularities such as Bloch points (BP) \cite{Feldtkeller1965} have been recently found to play a critical role in 3D magnetic textures hosting skyrmions \cite{Birch2020}  and hopfions \cite{Raftrey2021, Tejo2021}. 3D magnetic textures are of special interest in magnetic media because employment of 3D magnetic topological solitons like hopfions, domain walls (DW) and Bloch points (BP) can open a
\farkhad{a route to creating new kinds of magnetic information storage and spintronic devices based on 3D architectures.}


Ferromagnetic (FM) nanowires (NWs) are a simple system to confine and explore 3D magnetism that has attracted considerable attention upon the last decades. Cylindrical nanowires (having typical radii of 30-60 nm) made of ferromagnetic materials have been proposed as an alternative for the development of numerous electronic and spintronic devices. 
Nanowires have a robust microwave (\diego{MW}) absorption \cite{Dmytriiev2013}, whose frequencies can be adjusted within a wide range depending on the choice of FM material, which is ideal for the production of \diego{MW} devices 

One of the most studied \farkhad{ potential} applications of ferromagnetic NWs is the elaboration of magnetic memories. The possibility of arranging them in hexagonal arrays of high density \cite{Ivanov2016} makes them excellent candidates for this type of devices, since they allow reaching high information storage densities
Also, their shape allows the creation of vortex type flux-closure magnetic textures \cite{Ivanov2013, Andersen2020}, which reduce the magnetostatic interaction between the NWs. Specifically, the possibility of creating magnetic memories based on  DWs has been studied \cite{Wong2016, Weiser2010}, being inspired by the racetrack memory proposal \cite{Parkin2008} based on the storage of information in the form of DWs displaced by pulsed electrical currents.

Although the static properties of long cylindrical nanowires have been well studied in the recent years \cite{Ivanov2013, Gomez2018} with some indications on the possibility of magnetization reversal via BP singularity inside the vortex DW \cite{Bran2018}, investigation of the statics and \farkhad{specially} magnetization dynamics in the short NWs, where the DW and BP emerge, are absent.

\diego{BPs were first studied and named by Feldtkeller \cite{Feldtkeller1965}, consisting in exchange-dominated 3D magnetic textures that hide topological singularities with the defining property that around them every magnetization orientation is present only one time. BP-DWs, therefore, are complex 3D textures, composed by two facing magnetic vortices with head-to-head or tail-to-tail vortex cores that could normally appear in thick NWs, where the DW magnetization is allowed to curl around the NW axis, minimizing the magnetostatic energy \cite{Mermin1979}. }
\textit{To the best our knowledge, the BP-DW dynamics have been investigated only in relatively long (the wire length about 10 times of the diameter) NWs stabilizing the BP via bamboo-like shapes \cite{Berganza2016,Saez2022}.}

\begin{figure*}[t]
  \centering
  \includegraphics[width=0.75\linewidth]{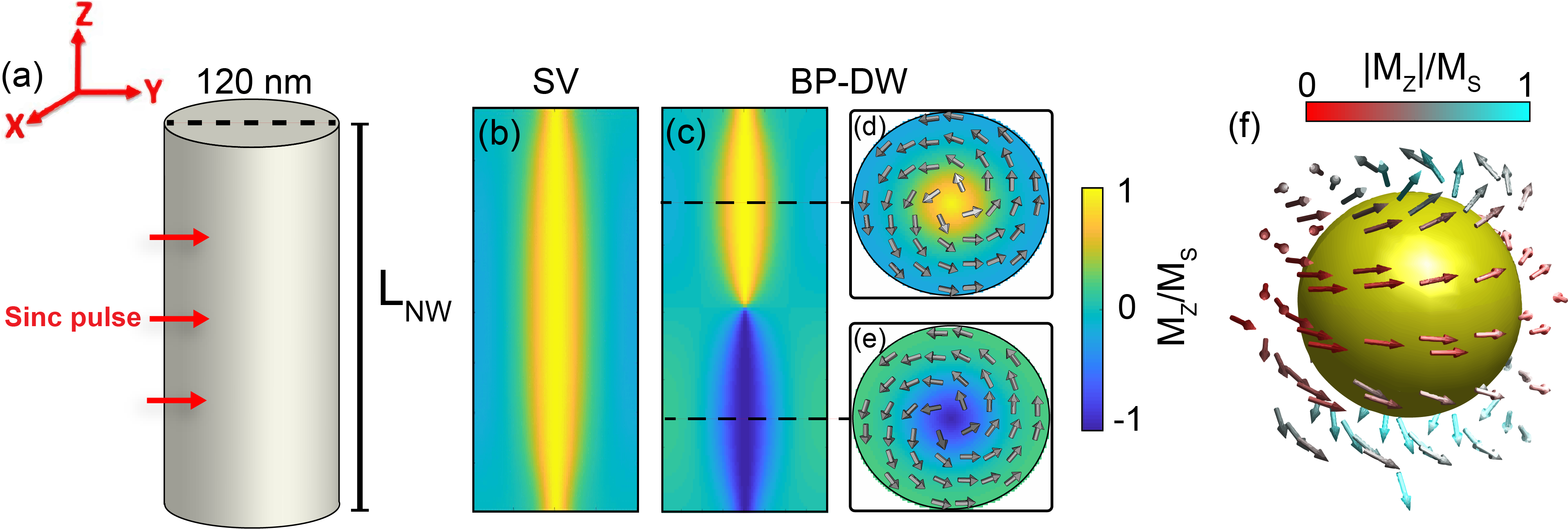} 
  \caption{
  Sketch of the cylindrical nanowire and its orientation with respect to the cartesian coordinate axes (a), showing the direction of the applied microwave magnetic field sinc pulse to excite the spin eigenmodes of the system. Vertical cut (xz-plane) of the 250 nm long nanowire showing a SV (b) and BP-DW magnetization configuration (c) with a vortex-like BP-DW in the middle. Arrows illustrate the in-plane magnetization and the colors are an indication of the out-of-plane magnetization. Transverse cut (xy-plane) of the nanowire in the BP-DW state are shown for the planes $z$ = 3L$_{NW}$/4 (d) and $z$ = L$_{NW}$/4, (e)  \diego{(see dashed lines in (c))}. Both vortex domains display opposite vortex polarities. (f) Three dimensional magnetization configuration surrounding the BP-DW centered in the middle a 250 nm long nanowire. Colors represent the magnitude of the axial-aligned magnetization.}
  \label{static}
  \end{figure*}

In this paper, through micromagnetic simulations and analytic theory, we carried out a detailed investigation of the BP-DW dynamics in relatively short (length up to three times of the diameter) cylindrical NWs, where the transition between the SV and vortex DW states confining the BP takes place. We also investigate the influence of the pinning on the stability of the BP. Finally, we suggest a \diego{device which uses MWs and SOT} to control of the transition between two different magnetic configurations, aiming at the creation of a topological magnetic memory based on short cylindrical NWs \farkhad {with fine frequency tuning and a latency time below the ns, i.e similar to one observed  with combined impact of SOT, STT and the voltage control of the magnetic anisotropy \cite{Grimaldi2020}}.


\section{Methods}
\subsection*{Simulation details}
\diego{Micromagnetic simulations were carried out in the open source software MuMax3 \cite{Mumax2014}}. The discretization cell size was set at 1.5 nm $\times$ 1.5 nm $\times$ 1.25 nm, \diego{much below the exchange length, which is approximately 3 nm for FeCoCu \cite{Berganza2016}, the material selected for our NWs.} 
The cell size is chosen based on the spacial scale of the magnetic topological textures which can appear in the system. We estimate a DW and vortex core thickness of about 10 nm. The NW diameter was set at 120 nm and its length is varied between 50 nm and 800 nm (statics) and 50 nm to 400 nm (dynamics). \farkhad{Previously studied NWs mainly were} 
either much longer (more than $1\, \mu m$ in  length) or \farkhad{were} almost disk-shaped (circular dots with the thickness much less than 100 nm). We used typical magnetic parameters of ferromagnetic \diego{FeCoCu} NWs in our simulations \cite{Bran2018}:
$$M_{s} = 2\,T ;\quad A_{ex} = 25\times 10^{-12}\, J/m ; \quad \alpha = 0.01 \quad ,$$where M$_s$ is the saturation magnetization of the NW.
and A$_{ex}$ and $\alpha$ describe the exchange stiffness constant and the damping of the NW respectively. $\alpha$ is reduced to 10$^{-4}$ in the dynamic simulations to better resolve the modes of the system.
For the purpose of simplicity and future reference in this study, we take the NWs axis parallel to the $z$ axis in the cartesian coordinate system (see Fig. \ref{static}a), i.e., the bases of the NW are placed parallel to the $xy$ plane.
Due to their cylindrical geometry, the NWs display a considerable shape anisotropy of magnetostatic origin, which favours the magnetization to be aligned with the NWs axis direction. No additional magnetocrystalline anisotropies were introduced in our simulations.

In order to uncover the main SW modes, the averaged in-plane magnetization of the system has been analyzed by means of a Fourier Transform \farkhad{\cite{Park2003}} to frequency domain after the application of a homogeneous magnetic field pulse in the following form:

\begin{equation}
\vec{H}_\text{pulse} = H_\text{pulse} sinc\left(\frac{2\pi t}{t_0}\right)\hat{u}_y ;
\end{equation}

where \diego{H$_\text{pulse}$} is the amplitude of the applied filed pulse, \diego{2 mT}, and t$_0$ is the full width at half maximum (FWHM) of the pulse, 1$\times$10$^{-12}$ s. The linearity of the excited SW modes has been checked up to the field magnitude of \diego{10 mT}, well above the used field amplitude. The function used for the pulse is sinc(x) = sin(x)/x, which has given better results than the Gaussian function, obtaining narrower and better defined peaks in the spin wave excitation spectrum given by the Fourier Transform. 

\begin{figure}[t]
  \centering
  \includegraphics[width=1\linewidth]{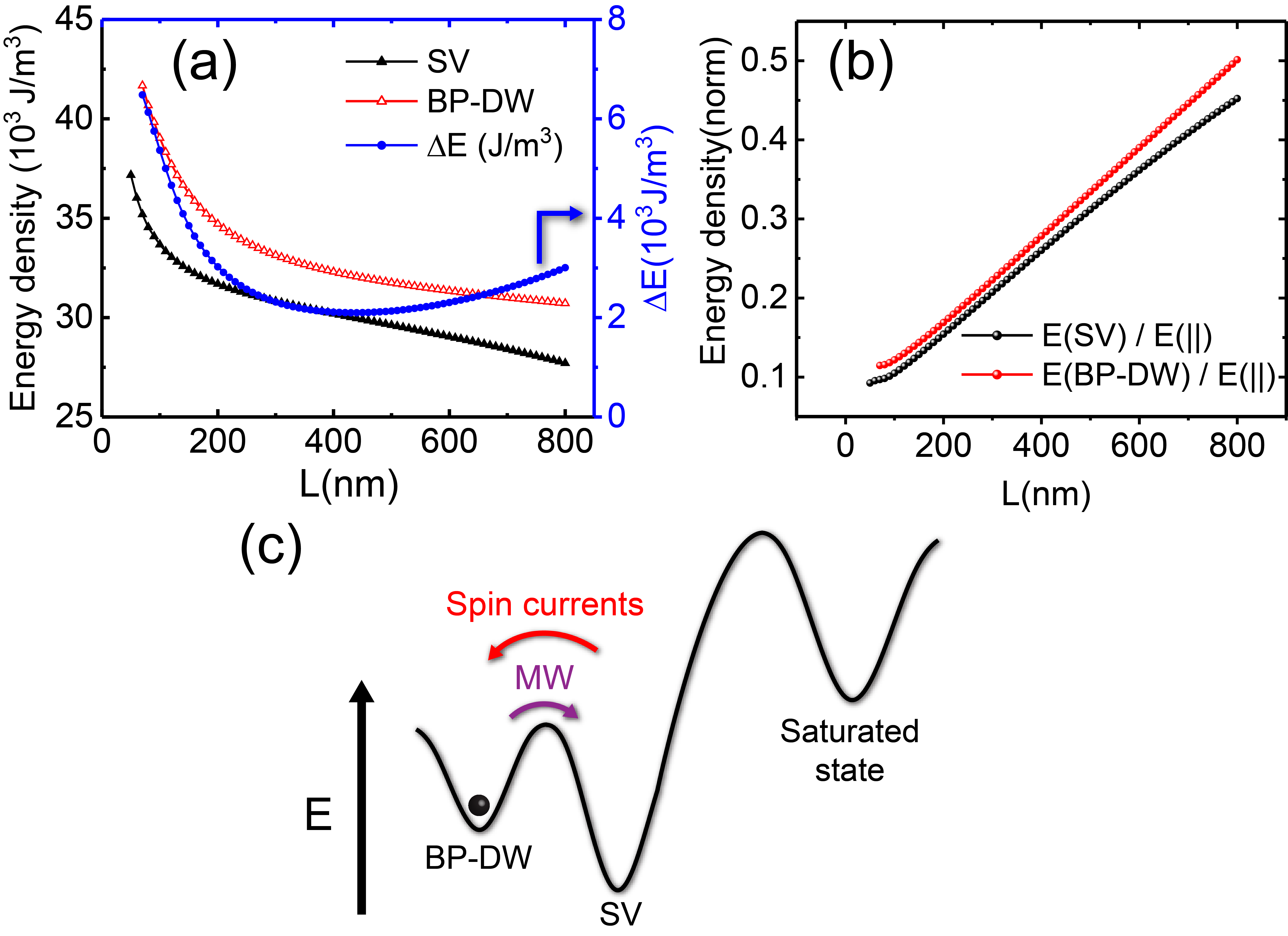} 
  \caption{\diego{(a) Total energy density of both studied magnetic configurations (SV and BP-DW) for NWs length comprehended between 70 and 800 nm and the gap ($\Delta$E = E(BP-DW) - E(SV)) between them in blue. (b) Energy density of the SV and BP-DW states for NWs length comprehended between 70 and 800 nm normalized by the parallel saturated state. (c) Sketch of the two level (BP-DW and SV) system and transitions that exist in the NWs of the studied length range (100-400 nm).}
  }
  \label{energies}
  \end{figure}
\section{Numerical results and discussion}

We analyze the simulation results in this section. First, the static magnetic configurations and related energies of NWs without \farkhad{induced} pinning are investigated as a function of the NW's length. 
Then, we investigate the main spin wave modes (in both SV and BP-DW states) and their evolution with the length of the NW. Finally, the ways of \diego{performing}
transitions between configurations with different topologies (i.e., SV vs. BP-DW) are presented.

\subsection{Static magnetization}

Depending on the NW's length and on the initial static magnetization configuration (ordered or disordered), the NW relaxation could potentially result in different stable states. If the final states correspond to a local minima of the total magnetic energy, it means that a relatively small excitation could make the system evolve into another neighbouring ground state. In other words, when a system  has topologically different but closely situated in energy metastable states, a relatively small excitation of the spin wave eigenmodes (involving small microwave power) may trigger the transition between those two states. We therefore introduce an analysis of the possible static magnetization configurations with its related energies.

 \begin{figure*}[t]
  \centering
  \includegraphics[width=0.6\linewidth]{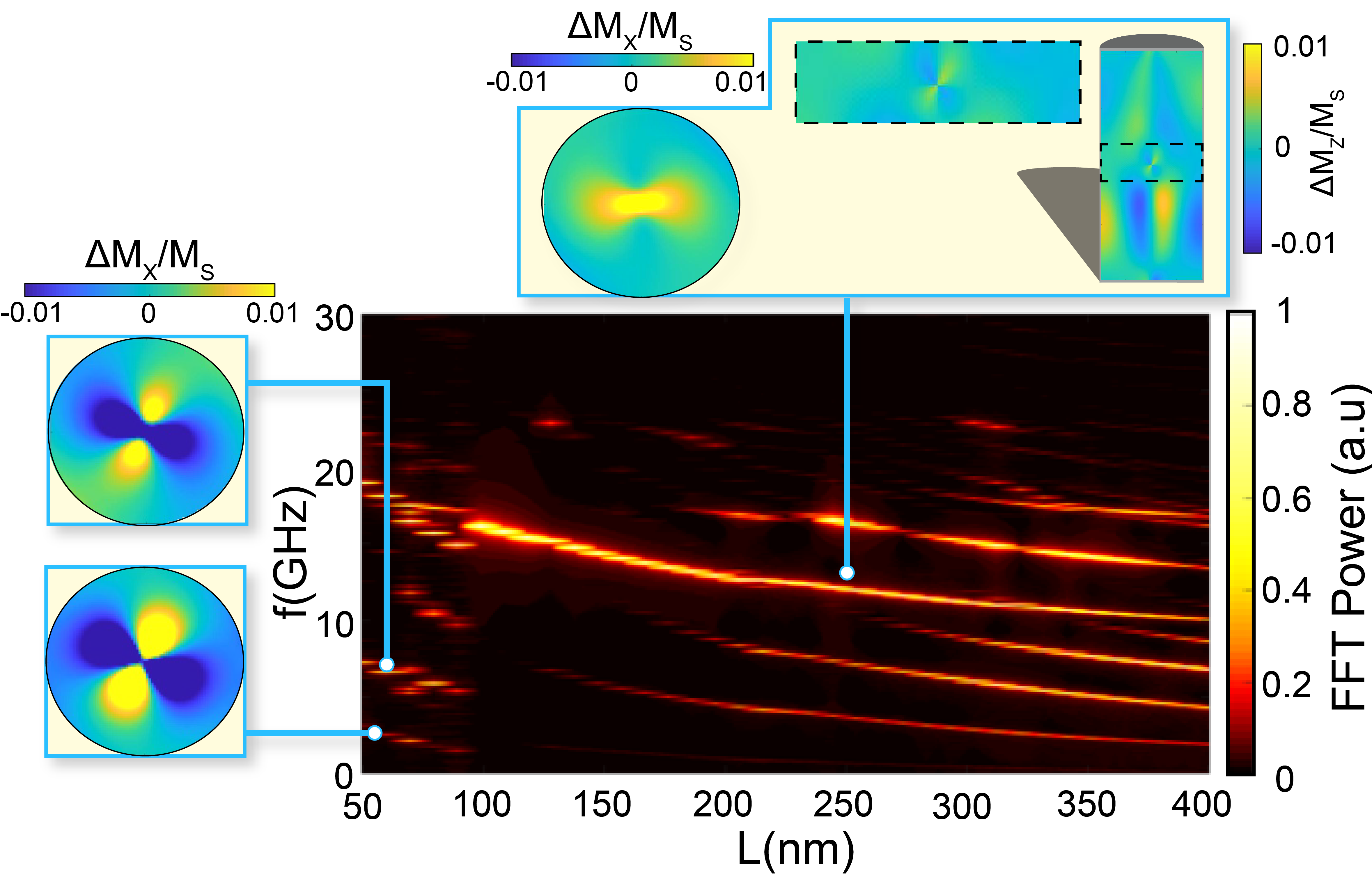} 
  \caption{Evolution of the main SW mode frequencies changing the nanowire length from 50 nm to 400 nm \diego{obtained after relaxing the BP-DW configuration. The frequency spectra for the NW lengths below 70 nm correspond to the SV state since the BP-DW state is not stable at such small NW aspect ratios}. The FFT power is normalized at each nanowire length. Snapshots of the eigenmode spatial distribution are shown for the main azimuthal modes in the SV state (the NW length is 60 nm) and the main mode at the length 250 nm at the nanowire top/bottom surface, as well as a longitudinal cross-section of the mode spatial distribution, focusing on the Bloch point vicinity (area delimited by the dashed line). }
  \label{modes_lenght}
  \end{figure*}

Fig. \ref{static}b shows the vortex state formed after relaxing a 250 nm long NW from a SV configuration, while Fig. \ref{static}c exhibits the BP-DW created after relaxing an initial configuration of two vortices with opposite polarities, one in each half of the nanowire (the front view of each vortex can be seen in Fig. \ref{static}d and Fig. \ref{static}e). \diego{Although the resulting BP-DW state is found to be stable (within reasonable simulation conditions of torque minimization \cite{Mumax2014}) it is well known that in practice, metastable states (such as for example the double vortex state, see Ref. \cite{Aliev2011}) are stabilised by the unavoidable natural pinning. 
We analyze the effect of the induced pinning on the stabilization of the BP-DW state in the Appendix \ref{appendix:pinning}.}

Our paper mainly concentrates on the intermediate length NWs (100-400 nm in length and with aspect ratio approximately between 1 and 3) whose static and dynamic response remained uncovered in the literature. Typical cylindrical nanowires or dots investigated before were either much longer ($\geq$ 1$\mu$m ) and multiple DWs \cite{Ivanov2016, Andersen2020} were created along their length when relaxed, or they remained shorter (length $\simeq$ radius), approaching a disk geometry where only the singular vortex state can be stabilized \cite{Aliev2009, Awad2010}. Fig. \ref{energies}a shows both magnetic configuration energy densities for NW lengths between 70 nm and 800 nm. Below 70 nm, the BP-DW state is not as stable, as it repeatedly falls into the SV state due to the magnetostatic energy of the NW, getting closer to resembling more the frame of a disk under those conditions. 

Each configuration energy has been found to be much lower in comparison to the magnetic energies of a saturated nanowire with \diego{a saturated magnetization parallel to the NW's axis (see Figure \ref{energies}b}, reinforcing the idea of a \diego{two energy-proximate state system which could reveal transitions from one to the other state with a given excitation}.

\diego{Figure \ref{energies}a} shows the difference between the energies of one (SV) and another (BP-DW) inhomogeneous magnetization states of interest ($\Delta$E = E(BP-DW) - E(SV)). The energy difference between these states is approximately equal to 5$\%$ of the total energy value. Therefore, these NW lengths seem to be adequate to make transitions between the single vortex and BP-DW states induced by external stimuli, since both states have similar energies (see Figure \ref{energies}c). This difference initially decreases with length, favoring the formation of DWs for longer NWs, and single vortex states for short ones. However, it starts increasing for NWs longer than 450-500 nm, revealing a minima in the range from 250 nm to 450 nm. Therefore, we can assume that in this length range the most efficient SV-BP transitions may occur. For the longest NWs, we have observed that the formation of the two BP-DWs states inside the NW reduces energy with respect to a single BP-DW state, explaining the origin of the relative increase of the energy of the BP-DW state for the longest NWs.

\subsection{Spin wave modes}

In this part we analyze numerically the main spin wave modes in SV and BP-DW states and study how the DW (and related BP) could be displaced under action of a \diego{MW} driving field.  In this study we take advantage of metastability of the BP-DW state in short NWs to explore the possibility to eliminate the BP-DW completely out of the NW via precessional magnetization motion by using a \diego{MW} excitation tuned to one of the most dominant spin wave mode (MM) frequencies.

\farkhad{As we shall see below, our method to manipulate the BP using MWs has \diego{arguments or considerations for and against in contrast} 
of the standard way to apply magnetic field. The main advantage of the method is the possibility to address the BP destruction individually by tuning the external frequency, which could be partially compensated by the need to apply opposite spin currents to restore the BP. Further below we suggest a practical implementation of the suggested method.}

Figure \ref{modes_lenght} shows the spin wave modes of a cylindrical NW for an initial BP-DW configuration for the NW lengths ranging from 50 to 400 nm. Even though the initial configuration is the BP-DW state, if the NW is short enough \diego{however, this will not be a stable configuration and} the system will relax into the SV state due to the magnetostatic energy contribution. \diego{This event is a direct consequence of the energy gap enhancement between the two states at shorter NWs lengths, which makes the system no longer metastable for lengths below 70 nm.}
This phenomenon is also translated to the dynamics of the system shown in Figure \ref{modes_lenght}: the detected modes for the NW length below 70 nm are different in nature from the ones above 70 nm. For the NWs that relax into a final BP-DW state, the frequency of the disclosed modes is progressively reduced with increasing the length of the NW. The SW modes are often being unfolded into new modes as the nanowire length increases, providing more complicated dynamics above about 260 nm in contra-position to the simplicity of the dynamics in the range of the NW length from 100 to 250 nm.

\begin{figure*}[t]
  \centering
  \includegraphics[width=0.8\linewidth]{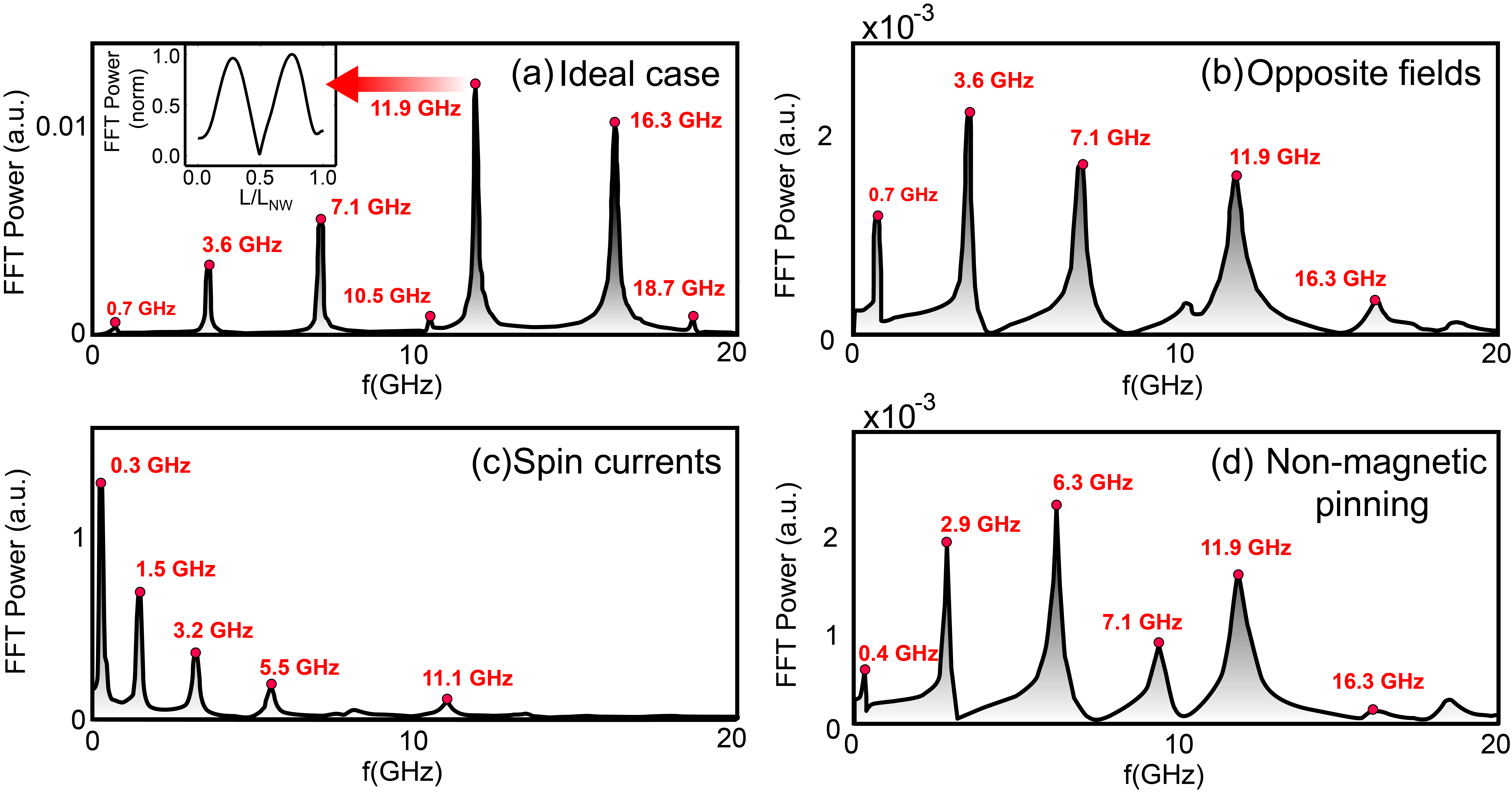} 
  \caption{Spin wave mode spectra of a 250 nm length NW in the BP-DW magnetization configuration originated from each of the four different methods explained in the text. \diego{Inset of (a) shows the normalized amplitude of the 11.9 GHz mode over the length of the NW.}} 
  \label{modes}
  \end{figure*}

Figure \ref{modes} shows the calculated spin wave modes in a BP-DW configuration of a 250 nm long NW after the BP-DW state was reached using different protocols. \diego{What are in our view the two foremost methods (the ideal case and the method of the spin currents) to achieve transitions between the two investigated states will be described in detail in Section \ref{subsec:Creation_BPDW}. For the opposite fields and non-magnetic pinning methods detailed explanations are in Appendices \ref{appendix:opp_fields} and \ref{appendix:BP-DW_created_by_pinning} and Appendix \ref{appendix:pinning} respectively.} 
Most of the main SW modes are presented in the majority of the reconstructed spectra. The mode located at 11.9 GHz stands out as the most \diego{prominent} when the BP-DW results from the ideal case. 
We have found that the difference between the spectra is fundamentally based on the position of the DW, which only in the ideal case (created by hand) turns out to coincide exactly with the center of the nanowire. \diego{Furthermore, the clear fronting and tailing of the modes observed in the spectras presented in Figure \ref{modes} (b, d) is a direct consequence of the asymmetric position of the BP along the NW.}

We have found that BP-DW states could be effectively transformed into SV state by \diego{MW} excitation tuned to the dominant SW mode (denominated here as BP-SV transition and analysed in details separately). Reverse (SV-BP) transition requires however more detailed investigation as energy of BP-DW state exceeds one of SV configuration. 


\subsection{Creation of the BP-DW state (SV-BP transition)}
\label{subsec:Creation_BPDW}
We turn now to investigate in details the SV-BP transition using as an example the 250 nm long NW, were the difference of the total magnetic energies between single vortex and DW configurations is close to minimum \diego{(see Figure \ref{energies}a)}.  
Achieving control of the SV-BP transition could the first step in getting a full control over the 3D magnetization textures and their topology in short ferromagnetic nanowires.

\subsubsection{BP-DW created by hand}
It was detailed above how a BP-DW could be generated in the middle plane of the nanowire with magnetization imposed as two perfect vortex states with opposite polarities followed by a relaxation. This is achievable with the precise control of the magnetization that \diego{the simulation} code grants but challenging to carry out \farkhad{in practice.} 
Due to the high symmetry that this system presents, with the BP-DW perfectly centered in the middle plane of the NW, we shall further refer it as the "ideal" case. Figure \ref{modes}a shows the corresponding frequency spectrum of the spin wave modes in the ideal case, presented with a dominant mode at 11.9 GHz and followed by a 16.3 GHz high-intensity mode. The 16.3 GHz mode seems to be relevant only in the ideal case and it has a much smaller amplitude or even does not appear at the rest of the (less symmetric) cases considered. \diego{A detailed analysis of the modes over the NW length reveals that these modes mainly excite the vortices itself, not the BP directly, which is what triggers the BP-DW motion (see inset of Figure \ref{modes}a showing the analysis of the 11.9 GHz mode).} 
See Ref.\cite{Videos} videos 1 and 2 showing the 11.9 and the 16.3 GHz modes excited in the ideal case.

\subsubsection{BP-DW created by spin current injection}

Having verified different ways to create the BP-DW configuration from the initial SV state, we shall further concentrate on, in our view, the most practical method which involves driving opposite spin currents into a ferromagnetic nanowire in order to impose the formation of the BP-DW state. 
In this section we show and describe the means to create DWs in 180-300 nm nanowires using spin currents.


\begin{figure}[t]
  \centering \includegraphics[width=1\linewidth]{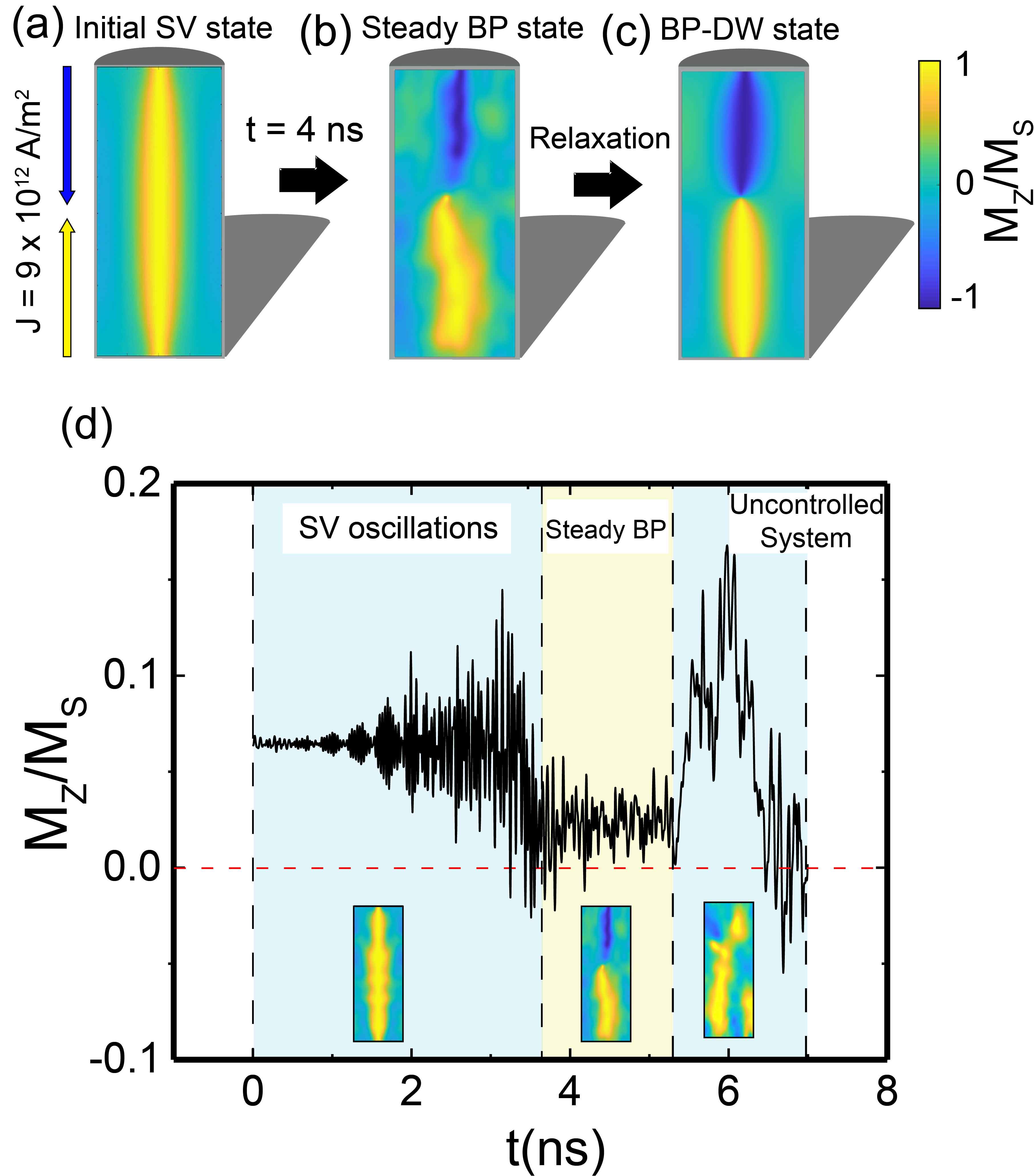} 
  \caption{Cross-section sequence of the m$_z$ magnetization component of the 250 nm long NW, showing the used method of 
  spin polarized currents to reconstruct the BP-DW state. 
  In \diego{(a)} a current of 9 $\times$10$^{12}$ A/m$^2$ is applied on each half of the NW, with opposite directions, to the initial SV state configuration. After 4 ns the steady BP state is reached \diego{(b)}. Switching off the currents and letting the system relax returns it into the DW state \diego{(c)}. \diego{(d)} Average axial-aligned magnetization component evolution in a 250 nm length NW with the spin polarized current density of 9 $\times$10$^{12}$ A/m$^2$. The three dynamic regimes are presented in the transition. Inset images represent a snapshot of the longitudinal cross-section of the axial $M_z$ component in the NW in each one of the regimes.}
  \label{spincurrents}
  \end{figure}

The application of spin polarized currents with a polarization degree P = 1 with opposite directions from each of the ends of the nanowire towards its interior, with an initial magnetic configuration of a single vortex \diego{(see Figure \ref{spincurrents}a)}, has been studied in details. The following procedure has been implemented in order to evaluate the minimum spin current density needed in order to successfully achieve a SV-BP transition: we leave spin currents simulations for long enough time (7 ns) to ensure the evolution of the system if the injected currents are strong enough. Then, the volume averaged m$_z$ component of the magnetization in the NW is evaluated over time, as it is an indicative of any transitions involving the polarity of the vortex core. In a typical spin current injection that promotes the system to transform from SV to BP-DW state \diego{(see Figure \ref{spincurrents}d)} there are three clear regimes: first, an oscillation of the SV magnetization induced by the injection of the spin currents. When the oscillations become large enough, the SV state \diego{is no longer manifested}. Two things can happen once this point is reached: if the currents are too strong, then the system does not evolve into a steady state but rather becomes uncontrollable, this is characterized by large and sharp ups and downs in the average m$_z$ magnetization. However, if the spin currents strength is precisely tuned, then there is an extra step in between that displays a steady BP-DW state, before turning again into an uncontrollable system \diego{(see Figure \ref{spincurrents}d)}. This state expresses itself as a durable, almost constant, close to zero average out-of-plane magnetization. \diego{See Ref.\cite{Videos} video 3 showing the spin current injection to a 250 nm NW to see the full procedure.}

\begin{figure*}[t]
  \centering
  \includegraphics[width=0.95\linewidth]{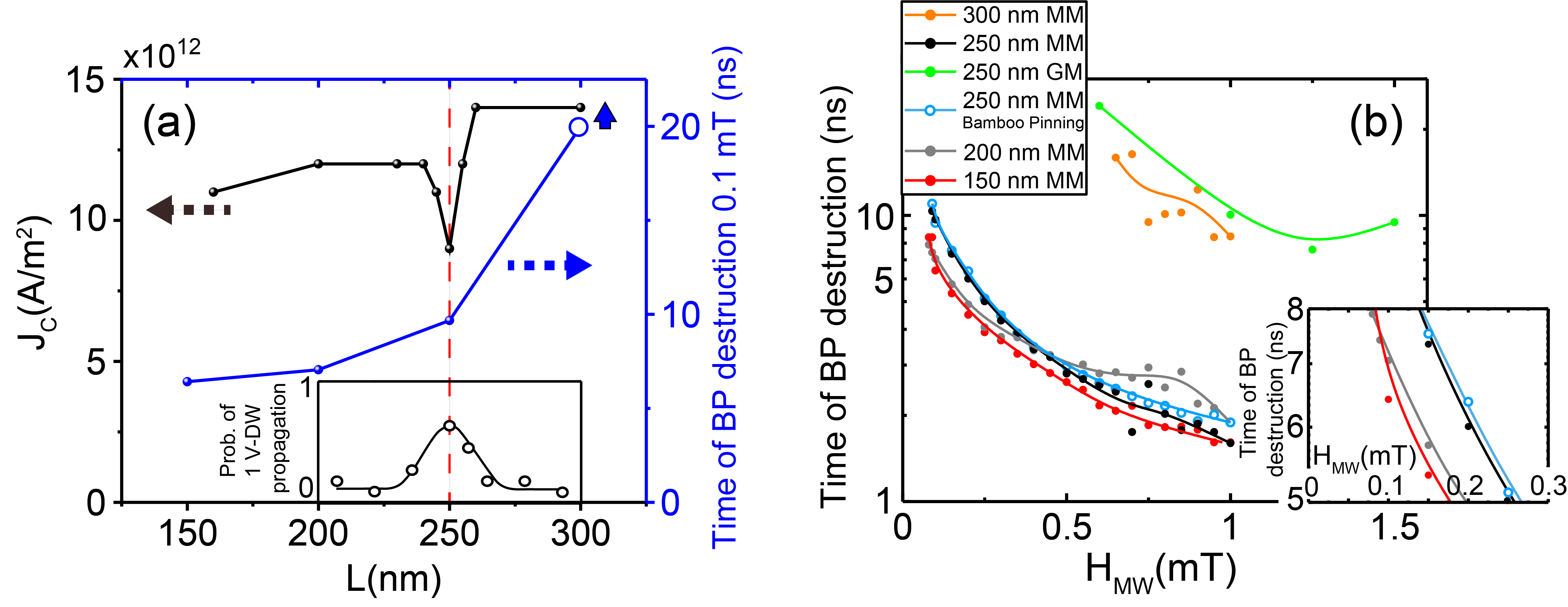} 
  \caption{ (a) Critical spin current density (J$_c$) needed to create a steady BP state and time of destruction of the BP-DW with the main mode at \diego{0.1 mT} plotted as a function of the NW length. \diegotwo{Inset of panel (a) depicts the probability to detect a 1 V-DW propagation reversal mode against the NW length.}(b) Time required for the \diego{MW} excitation to start to displace the DW, in logarithmic scale, plotted against the excitation field amplitude, for different lengths of the NW. For the NW length of 250 nm, the main mode (MM) and gyrotropic mode (GM) have been studied. Inset corresponds to a zoom into the main mode excitations for the NW length ranging from 150 to 250 nm. The point at 300 nm in the time of destruction in (a) is a lower end estimation following its trend on (b). }
  \label{tiempo_destruct}
  \end{figure*}
  
The protocol shown in Figure \diego{\ref{spincurrents}(a, b, c)} for obtaining BP-DW by using spin currents is as follows: first, we apply opposite spin currents from each NW's end towards its interior, then the currents are removed once the steady BP-DW state is reached \diego{(see Figure \ref{spincurrents}b)}. By letting the nanowire relax from the steady BP-DW state, the desired BP-DW state could be obtained. Figure \diego{\ref{spincurrents}b} shows the magnetization  of the nanowire  after the application of spin currents of  9 $\times$ 10$^{12}$\, A/m$^2$ for 4 ns, from which state the NW is relaxed to a BP-DW configuration.

Using this protocol and the appropriate current intensities, it is indeed possible to reconstruct the desired BP-DW state. 
For all studied NW lengths the spin currents needed to generate the BP-DW were within the range  $9\times10^{12}$  A/m$^2$ and $14\times10^{12}~\text{A/m}^2$ (see Figure \ref{tiempo_destruct}a). Interestingly, close to L=250 nm 
a minimum in the critical spin current density $J_c$ required to obtain BP-DW state emerges.

Our numerical experiments revealed that the efficiency of the spin current induced reversal SV-BP transition substantially increases in the vicinity of NW length $L$ = 250 nm (see drop of the minimum spin current needed shown in Figure \ref{tiempo_destruct}a). In order to figure out the possible origin on this unexpected behaviour, we have investigated the variation of the magnetization reversal modes as a function of the NW length. 

\diegotwo{Previous studies of the phase diagram of the magnetization reversal modes in cylindrical NWs showed that the reversal modes are indeed strongly dependent on the NW size \cite{Proenca2021}. Their aspect ratio determines the type of DW formed (transverse, T-DWs, or vortex DW, V-DWs) and the number of nucleated DWs \cite{Proenca2021}. Although the saturation magnetization used in our work is higher than one used in the work by Proenca et al. \cite{Proenca2021}, the expected limits of the transverse to vortex DWs mediated reversal modes transition should, in principle, appear for lower nanowire diameters than ours. Indeed, for NWs with a diameter of 120 nm investigated in this work, our simulations confirm that all magnetization reversal modes are V-DWs. However, the reversal modes are quite sensitive to the length of the NWs, providing variations on the presence of the different observed reversal modes even for the limited range of the studied NW lengths (see inset to Figure \ref{tiempo_destruct}a).}

\diegotwo{In particular, for the length range between 220 nm and 280 nm, we observed two possible reversal modes in our simulations, related to a spiral rotation of two V-DWs (2 V-DW) or, in other cases, the propagation of one V-DW (1 V-DW). Interestingly, the maximum in the probability of the 1 V-DW propagation magnetization reversal coincides with the maximum efficiency in the spin current injection (located at a nanowire length 250 nm, see Fig. \ref{tiempo_destruct}a). We attribute this effect to a generally higher vortex symmetry in this NW length range which matches to the highly symmetric spin current injection implemented to accomplish the backward SV-BP transition. Details of the estimation of the probability of the V-DW mediated magnetization reversal (depicted in the inset of Figure \ref{tiempo_destruct}a) are provided in Appendix \ref{appendix:reversalmodes}.}


\subsection{Optimization of the control over magnetization topology}


In the previous sections SW modes in nanowires for the BP-DW configurations 
have been disclosed (Figures \ref{modes_lenght}, \ref{modes}). One could expect that these configurations under a \diego{MW} excitation at a frequency close to one of the main eigenfrequencies of the nanowire would ultimately result in a transition between the BP-DW (a metastable state) and the SV state (a stable state). To verify this hypothesis, a sinusoidal excitation magnetic field has been applied to each magnetization configuration as follows: $\vec{H}_\text{excitation} = H_\text{MW}\sin(2\pi ft)\hat{u}_y$. Where  H$_\text{MW}$ is the excitation amplitude and 
$f$ is the frequency of the SW mode used to excite the NW. The \diego{MW} field applied throughout the whole NW has been directed perpendicularly to the NW axis, similarly to the sinc pulse used before to disclose the SW eigenmodes of the system (see Figure \ref{static}a).

The result of applying the \diego{MW} field is as follows: 
after a certain time (which depended on the intensity of the excitation and the used SW mode frequency), the BP-DW moves towards one of the two NW's ends and the SV state is created in the NW (see Ref.\cite{Videos} video 4 showing the BP-DW state destruction with the application of a \diego{MW} excitation field with the frequency of the main SW mode). Figure \ref{tiempo_destruct}b shows dependence of the time of application of the \diego{MW} microwave excitation required to remove BP-DW as a function of the excitation amplitude for different NW lengths. The 11.9 GHz mode has been chosen for this study as the main SW mode (MM), since it is notably present in all SW excitation spectra shown in Figure \ref{modes}. The time of destruction becomes effectively larger if the excited SW mode is not the main in amplitude. For example, if the 250 nm BP-DW nanowire is excited with a frequency close to its gyrotropic mode frequency (GM, located at 0.69 GHz for 250 nm NW length), which is directly related to the BP position oscillations and should induce a more localized excitation of this 
magnetization texture, then the time of BP-DW destruction becomes much larger (see Figure \ref{tiempo_destruct}b), even though, over time, the transition is accomplished. \diego{Latency times below the ns are accomplished in our system by exciting the most predominant SW mode (11.9 GHz) with MW fields exceeding \diego{5 mT} \diegotwo{(see Appendix \ref{appendix:latency})}.} 

Figure \ref{tiempo_destruct}b shows that the BP-DW to SV state transition takes longer to occur as the amplitude of excitation decreases. For \diego{MW} excitation amplitudes smaller than those shown, the displacement of the DW was not observed for times below 10 ns. It can be seen that the transition occurs for relatively short times even with small \diego{MW} field amplitudes ($ \approx \ $ \diego{0.2 mT} ). Figure \ref{tiempo_destruct}b also shows a big leap in the efficiency of the BP-DW destruction changing the NW length from 250 nm to 300 nm. This effect may be attributed to a transition from a regime in which several SW modes of similar amplitude are involved in the dynamics of the system, 300 nm long NWs, as opposed to a regime in which the dynamics relay on one dominant main mode, 250 nm and shorter NWs (see Figure \ref{modes_lenght}).  
This recipe, therefore, presents a way of moving the domain walls in a nanowire with relatively small energy cost. The BP velocities in the destruction process reach $\approx$ 400 m/s, demonstrating such high velocities even with a small applied field.
 
Figure \ref{tiempo_destruct}a compares the time required to remove the BP-DW using a \diego{MW} excitation of \diego{0.1 mT} of the main SW mode and the minimum spin current required to restore BP-DW in the NW length range, where such reversible transitions were found to be possible to perform. The energy required for the BP-DW destruction-and-restoration cycle can be obtained from these two graphs. Simulations suggest that the needed time for the BP-DW destruction notably increases in the NW length range 250-300 nm due to 
topological differences arising from the fact that a single BP-DW is more stable at lengths in this mentioned range (see Figure \ref{tiempo_destruct}b). However, for the NW lengths from 150 to 250 nm the time of destruction is almost maintained. The reconstruction with the spin currents is most effective at the NW length of 250 nm, for which there is a minimum of the spin current density $J_c$ necessary to reconstruct the BP-DW. We conclude that the optimal length for the SV to BP-DW cycle efficiency is close to 250 nm. 

\begin{figure}[t]
  \centering
  \includegraphics[width=0.8\linewidth]{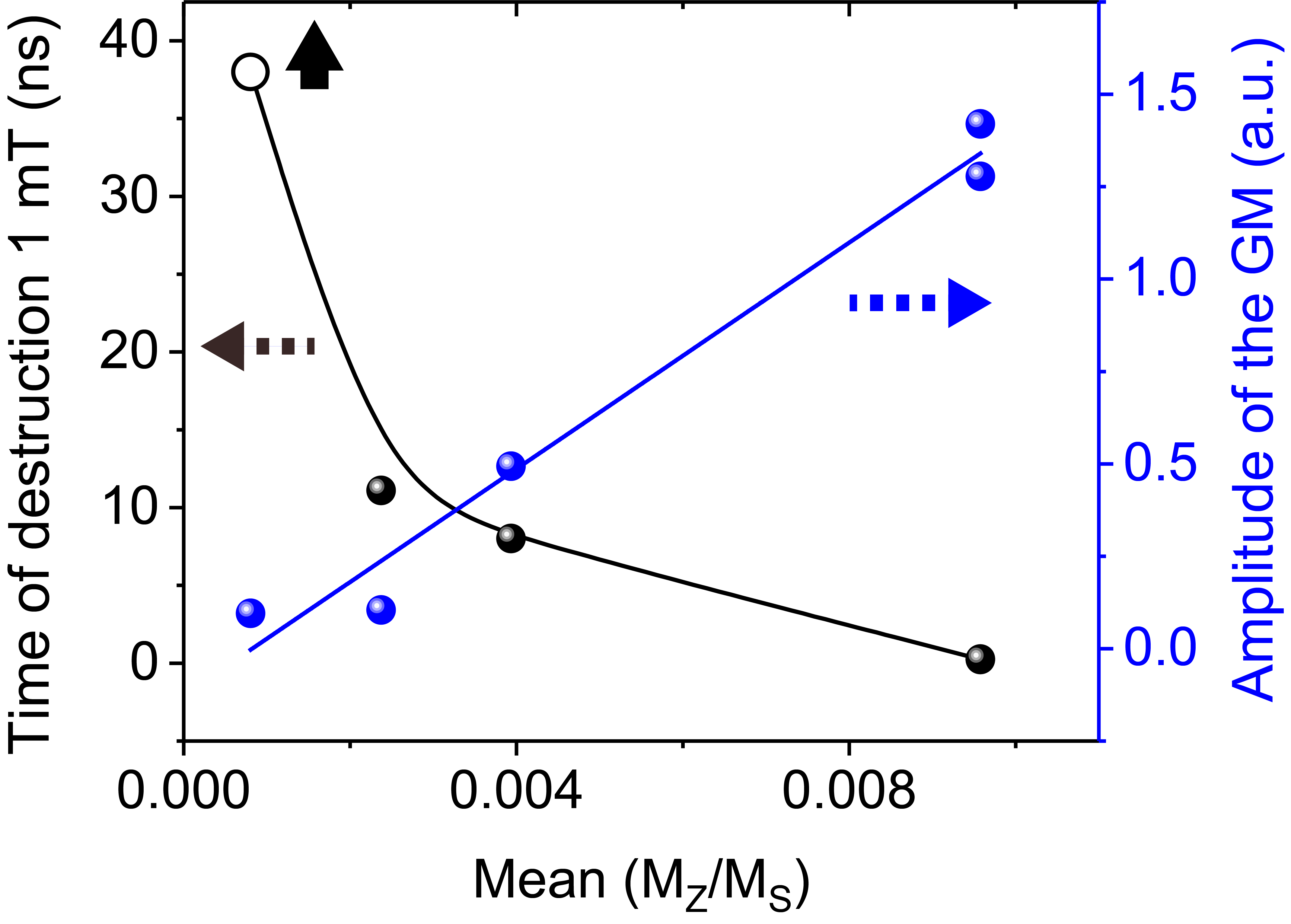} 
  \caption{Time of destruction of the BP-DW state with an \diego{MW} excitation field with the amplitude of \diego{1 mT} and the frequency of GM and corresponding amplitude of the GM for different stopping times after the application of a 9 $\times$ 10$^{12}$ A/m$^2$ spin current density in a 250 nm nanowire plotted against the average out-of-plane magnetization component. The point at almost 40 ns is a lower estimation of the destruction time.}
  \label{gyrotropic_mode}
  \end{figure}
 
\subsection{Optimization of the control over the magnetization topology using the gyrotropic mode}

Up until this point, the destruction of the BP-DW has been performed administrating a small \diego{MW} excitation with the frequency of the detected main SW mode of 11.9 GHz. The magnitude of this mode in comparison with the others allows the destruction to be efficient both in time and in required amplitude of the applied \diego{MW} field (see Figure \ref{tiempo_destruct}b). However, the mode spectra generated after the reconstruction of the BP-DW via spin currents displays a redistribution of the modes amplitudes. Once the BP-DW state is created by injection of opposite spin currents, the gyrotropic mode becomes also the main eigenmode of the system  (see Figure \ref{modes}c), meaning that it might be now more efficient to destroy the new BP-DW state by using the gyrotropic mode frequency as the \diego{MW} excitation frequency.

Interestingly, the obtained results indicate that the effectiveness of using gyrotropic mode to realize  the BP-DW to SV state transition strongly depends on the BP-DW symmetry with respect to center of the nanowire. As mentioned above, to obtain the BP-DW, the opposite spin current injection should be stopped in the steady BP-DW regime shown in Figure \diego{\ref{spincurrents}d}. However, even in this regime, the BP-DW does not completely resemble the configuration obtained in the ideal case, but it is rather dependent on the exact moment of relaxation to result on a more or less symmetric outcome. As Figure \ref{gyrotropic_mode} shows, the time of destruction of the spin current reinstated BP-DW state decays as the system becomes more asymmetric. Simultaneously, the amplitude of the GM excited to accomplish the BP-DW to SV state transition increases with the asymmetry of the BP-DW position with respect to the center of the nanowire.

\section{Analytical model for dynamics of the Bloch point}



In order to provide analytical estimation for the lowest (gyrotropic) spin wave frequency in a nanowire we consider a BP-DW in a cylindrical nanowire with a length $L$ and radius $R$. We use the cylindrical coordinate system ($\rho$, $\phi$, z) with the O$z$ axis coinciding with the nanowire axis. We assume that the domain wall is a vortex type and the $\rho$-magnetization component is equal to zero, $m_\rho$ = 0. The $m_z$ magnetization component we choose in the azimuthally symmetric trial form suggested by Thiaville et al. \cite{Thiaville2003}
\begin{equation}
m_z(\rho,z) = \frac{z}{\sqrt{\rho^2 + z^2}}\frac{b^2}{b^2 + \rho^2}
\label{mz}
\end{equation}
 where $b$ is the BP size (order of 10 nm).
To calculate the BP excitation gyrotropic frequency, we use Landau-Lifshitz-Gilbert (LLG) equation of motion, which describes the evolution of magnetization \textbf{M}(r,t) in a ferromagnet in the presence of an effective magnetic field \textbf{H}$_{eff}$ and damping. It is written in the following form: 
\begin{equation}
\frac{d\textbf{M}}{dt} = -\gamma\textbf{M}\times\textbf{H}_{eff} + \frac{\alpha}{M_s}\textbf{M}\times\frac{d\textbf{M}}{dt},
\label{LLG}
\end{equation}
where $\gamma$ is the gyromagnetic ratio, $\alpha$ is the dimensionless damping constant and $M_s$ is the saturation magnetization. The effective field \textbf{H}$_{eff}$ = -$\delta$E/$\delta$\textbf{M} is a combination of the external magnetic field, exchange field and the demagnetizing field, where E is the total magnetic energy density. It was shown by Thiele \cite{Thiele1973} that for description of the dynamics of magnetic domains the LLG equation can be rewritten in another form that allows simplifying the calculations. The Thiele’s approach is applicable to describe dynamics of stable magnetization configurations (magnetic solitons) that can be characterized by a position of its center \textbf{X}(t) (a collective coordinate) that can vary with time. To consider BP gyrotropic dynamics we assume that magnetization as a function of the coordinates \textbf{r} = ($\rho$, z) and time can be written in the form \textbf{M}(\textbf{r},t) = \textbf{M}(\textbf{r}, \textbf{X}(t)). The three-dimensional vector \textbf{X}(t) represents oscillations of the Bloch point center position. Then, neglecting damping, we can rewrite the LLG equation as:
\begin{equation}
G_{\alpha\beta}\frac{dX_\beta}{dt}=-\frac{\delta W}{\delta X_\alpha},
\label{LLG-rewritten}
\end{equation}
where W is the total magnetic energy, $\alpha$, $\beta$ = $x, y, z$. Introducing the unit vector of magnetization \textbf{m}=\textbf{M}/M$_s$, the components of the global (volume averaged) gyrotensor G, can be defined as follows:
\begin{equation}
G_{\alpha\beta} = \frac{M_s}{\gamma}\int d^3\textbf{r}\left(\frac{\partial \textbf{m}}{\partial X_\alpha}\times\frac{\partial \textbf{m} }{\partial X_\beta}\right)\textbf{m}
\label{gyrotensor}
\end{equation}

The dependence \textbf{M}(\textbf{r},\textbf{X}(t)) is unknown, and, therefore, we make the simplest assumption \textbf{M}(\textbf{r},\textbf{X}(t)) = \textbf{M}(\textbf{r} - \textbf{X}(t)), that corresponds to the BP rigid motion around the equilibrium position \textbf{X} = 0. We assume that metastable equilibrium BP position is located on the nanowire axis $Oz$ at the point $z=0$. The corresponding vortex type DW containing this BP is located in the $z=0$ plane separating the nanowire in two symmetrical domains with magnetization directed upwards ($z>0$) and downwards ($z<0$). 
The global gyrovector given by Eq. (\ref{gyrotensor}) can be calculated at the BP position \textbf{X} = 0 using the standard approach as

\begin{equation}
G_{\alpha\beta} = \frac{M_s}{\gamma}\int d^3\textbf{r}F_{\alpha\beta}^{e},
\label{gyrovector}
\end{equation}

where F$_{\alpha\beta}^{e}$ is the emergent electromagnetic field tensor F$_{\alpha\beta}^{e}$  = $\left(\frac{\partial \textbf{m}}{\partial X_\alpha}\times\frac{\partial \textbf{m}}{\partial X_\beta}\right) \textbf{m}$ \cite{Guslienko2016}.

The tensor G$_{\alpha\beta}$ is antisymmetric and, therefore, it can be represented via a dual vector \textbf{G}, known as gyrovector. Substituting the magnetization distribution (\ref{mz}) to Eq. (\ref{gyrovector}) one can calculate the global gyrovector components in the cylindrical coordinates as \textbf{G} = ($G_{\rho}$, 0, 0). The $\rho$-component for the case $L >> R$ is  

\begin{equation}
G_\rho = -4\pi b \frac{M_s}{\gamma}atan\left(\frac{R}{b}\right).
\label{gyrovector-cil}
\end{equation}

The global gyrovector is finite and is a function of the BP radius $b$. The equilibrium value of $b$ can be found from minimization of the total magnetic energy of centered BP consisting of the exchange and magnetostatic energies. 
We note that that the nullification of the global gyrovector is a property of infinite systems with specific asymptotic behavior of the magnetization at the infinity ($\textbf{m}$=const). Such magnetization textures are localized 3D solitons, toroidal hopfions, for instance \cite{Tejo2021}.
The Bloch point is a non-localized 3D magnetic topological soliton and may reveal a non-zero global gyrovector in restricted cylindrical NW geometry given by Eq. (\ref{gyrovector-cil}). The finite global gyrovector allows to calculate the BP gyrotropic frequency. The non-zero transverse ($G_\rho$) gyrovector component leads to expelling of the BP to the NW circumference at any value of the BP-DW velocity directed along the NW length due to the gyroforce. 
The SV state magnetization does not depend on the thickness coordinate $z$ except some small area near the nanowire faces. Therefore, SV state is essentially two-dimensional and described by 2D magnetic topological charge, which is proportional to the $z$ component of the vortex gyrovector, $G_z = -2\pi L \frac{M_s}{\gamma}$. The global gyrovector of the BP-DW is a result of more complicated 3D magnetization distribution given by Eq. (\ref{mz}). The ratio of the BP and SV gyrovector absolute values, $\frac{G_\rho}{G_z}= \frac{\pi b}{L}$, is small for the nanowire lengths 100 - 300 nm investigated in the present article.

Accounting for the azimuthal symmetry of the system the moving BP energy for small BP position oscillations near the equilibrium in the NW center can be written as:
\begin{equation}
W(\textbf(X)) = W(0) + \frac{1}{2}\kappa (X^2 + Y^2) + \frac{1}{2}\kappa_z Z^2
\end{equation}

Solving the Thiele equation of motion (\ref{LLG-rewritten}) for the BP center position \textbf{X}(t) one can find the gyrotropic oscillation frequency

\begin{equation}
\omega_G = \frac{\sqrt{\kappa \kappa_z}}{G_\rho}.
\label{frequency-gyr}
\end{equation}

The stiffness coefficients $\kappa$, $\kappa_z$ were calculated numerically, see Table I. 
The equilibrium value of $b$ is 7.2 nm for the given nanowire parameters. Using the data from Table I and Eq. (\ref{frequency-gyr}) one can calculate the BP gyrotropic frequency to be $\omega_G$/2$\pi$ = 0.45 GHz without pinning and $\omega_G$/2$\pi$ = 2.20 GHz with pinning.

\begin{table}[t]
\begin{center}
\begin{tabular}{| r | l | c |}\hline
 & \textbf{Without pinning} & \textbf{With pinning} \\ \hline
\textbf{Transverse disp.} & 1.2336$\times$10$^{-22}$ & 2.67208$\times$10$^{-22}$ \\ \hline
\textbf{Axial disp.} & 8.08524$\times$10$^{-20}$ & 8.76542$\times$10$^{-19}$ \\ \hline
\end{tabular}
\caption{Second derivative of the BP energy (in units of J/nm$^2$) with respect to the BP's displacements under applied field in transverse and axial directions in the nanowire. The simulated nanowire has 120 nm of diameter and 250 nm of length. Used magnetic parameters are: $\mu_0 M_s$ (saturation magnetization) = 2 T, $A_{ex}$ (exchange stiffness constant) = 25 pJ/m and $\alpha$ (damping) = 0.01. Pinning is made by increasing two times the saturation magnetization in the middle of the NW where BP is located. }
\label{tab:derivative}
\end{center}
\end{table}

\begin{figure}[ht]
  \centering
  \includegraphics[width=1\linewidth]{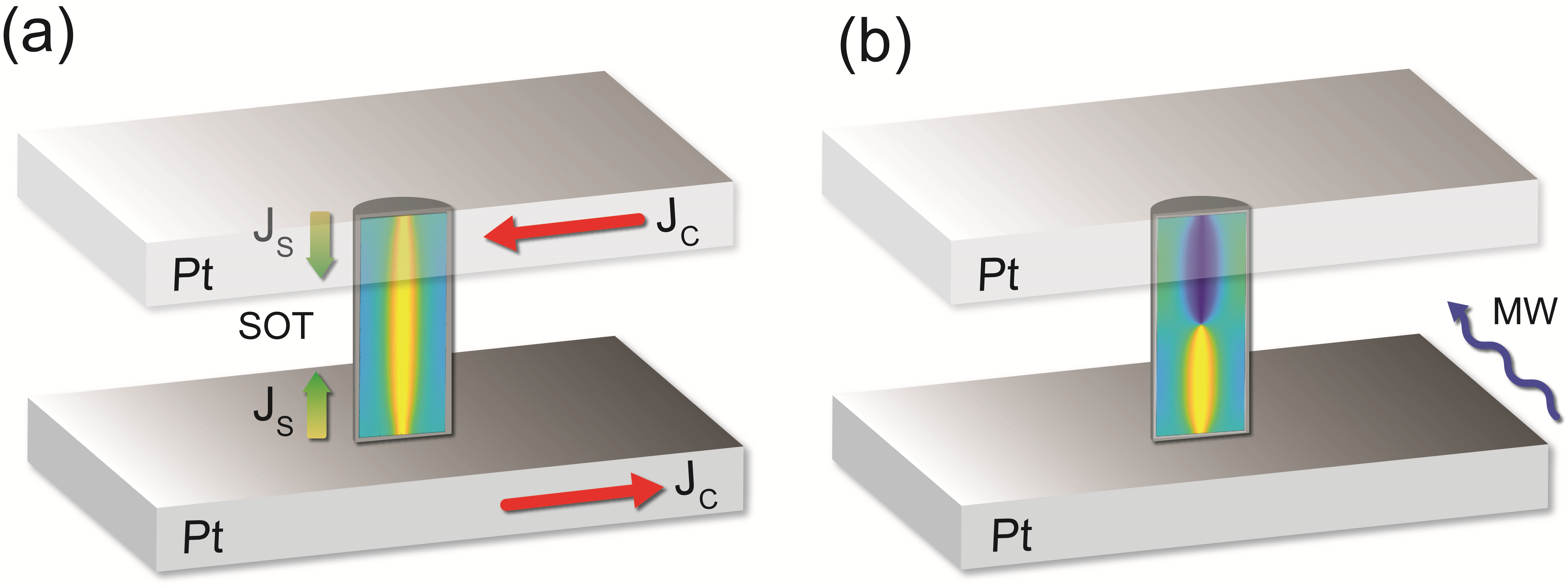} 
  \caption{\diego{Schematic of the suggested device to implement the described transition mechanism. (a) Electric current is simultaneously applied through two Pt electrodes contacting the opposite NW ends in the SV state, triggering two opposite spin polarized currents through NW by SOT, effectively switching the NW to the BP-DW state. (b) MW excitation is applied perpendicularly to the NW in the BP-DW state to trigger the transition into the SV state.
}}
  \label{device}
\end{figure}

\diego{\section{Device implementation}}
\label{subsec:device}

\diego{A suitable device to implement the proposed reversible BP-DW manipulations could include two heavy metal (Pt) electrodes located on the ends of the FeCoCu NW. These electrodes would be able to generate the spin polarized current via SOT once electric current through Pt is applied, effectively restoring the BP-DW state (see Figure \ref{device}a). We note that the SOT is known to provide much higher energy efficiency and lower latency times than STT \cite{Saha2022}. Additionally, the device would avoid possible heating in the NW since the spin current and charge current are independent separated events, which would not happen with STT.}

\diego{For the BP-SV transition (see Figure \ref{device}b), external frequency-tuned MW is needed, one way to achieve this is placing the sample on a coplanar waveguide (CPW) and apply magnetic pulses. Another way is to take advantage of the two Pt electrodes and use them as a waveguide. In the latter case the MW excitation will have a predominant axial direction. This is however is not an issue since we have checked that the BP-SV transition can indeed be also achieved using an axial excitation and with similar latency times as in the case of a perpendicular MW excitation.}
\\
\section{Conclusions}

In summary, we have investigated numerically and by analytical theory static and dynamic (spin wave modes) properties in short cylindrical magnetic nanowires, where the ground (single vortex) and metastable (vortex domain wall with a Bloch point) states have similar energies. We have demonstrated effective reverse control over the 3D magnetization texture topology switched between these two states by using a low power microwave short microwave excitation. To destroy the BP-DW state the \diego{MW} field frequency was chosen to be close to one of the dominant SW mode frequencies. The opposite spin polarized currents were used to reestablish the BP-DW magnetization configuration. \diego{Finally, an implementation of the mechanism has been proposed on a practical device involving the generation of the opposite spin currents via SOT}. Our results path a way towards the creation of a novel type of energy efficient magnetic memories based on short ferromagnetic cylindrical nanowires.


\section{acknowledgments}
The work in Madrid was supported by Spanish Ministry of Science and Innovation (RTI2018-095303-B-C55, PID2021-124585NB-C32, TED2021-130196B-C22) and Consejería de Educación e Investigación de la Comunidad de Madrid (NANOMAGCOST-CM Ref. P2018/NMT-4321) grants. F.G.A. acknowledges financial support from the Spanish Ministry of Science and Innovation, through the María de Maeztu Programme for Units of Excellence in R\&D (CEX2018-000805-M) and "Acción financiada por la Comunidad de Madrid en el marco del convenio plurianual con la Universidad Autónoma de Madrid en Linea 3: Excelencia para el Profesorado Universitario". J.C. thanks Spanish MECD for the fellowship. K.G. acknowledges support by IKERBASQUE (the Basque Foundation for Science). K.G. work was supported by the Spanish Ministry of Science and Innovation under grant PID2019-108075RB-C33/AEI/1013039/501100011033 and by the Norwegian Financial Mechanism 2014–2021 trough project UMO-2020/37/K/ST3/02450.

\clearpage
\section*{Appendix}
\appendix

\section{BP-DW created by opposite magnetic fields}
\label{appendix:opp_fields}
The first proposed method that could be experimentally achievable consists of applying two homogeneous magnetic fields in the direction of the NW axis to the vortex state, with opposite directions in each of the two halves (further called opposite magnetic fields method). If the nanowire has its main axis in the direction of the Z axis and its center is at the origin of coordinates, we would apply the following magnetic field:

\begin{equation}
\vec{H} = \left\{\begin{matrix}
H\hat{u}_z\quad \,\,\,\, z<0\\ 
-H\hat{u}_z\quad z>0
\end{matrix}\right.
\end{equation}

The protocol to follow with this method is as follows: (i) apply the magnetic field $\vec{H}$ to the nanowire (ii) relax the nanowire with the applied field (iii) remove the field and relax the nanowire.

\begin{figure}[ht]
  \centering
  \includegraphics[width=1.0\linewidth]{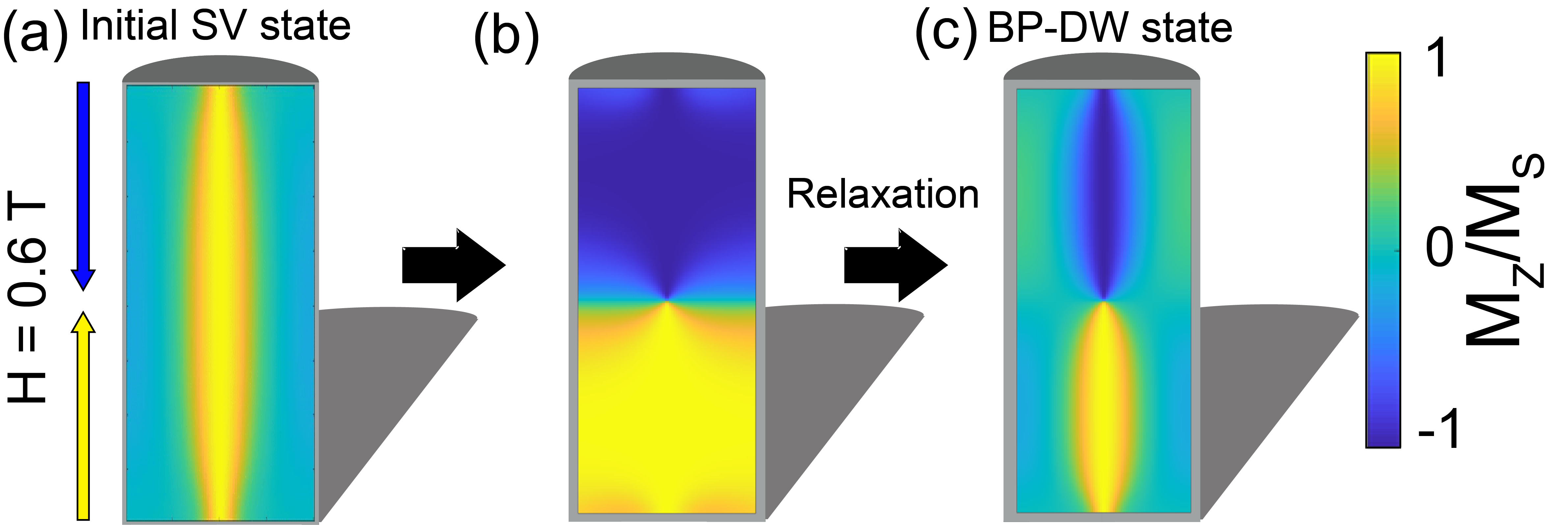} 
  \caption{ Cross-section sequence of the m$_z$ magnetization component of the 250 nm long NW, showing the used method of opposite magnetic fields to reconstruct the BP-DW state. (a) Displays the initial SV state before the opposite magnetic fields applied. After two \diego{0.6 T} opposite fields, the m$_z$ magnetization component is almost completely saturated on each half of the NW (b). After relaxing the NW the BP-DW is reached (c). }
  \label{opposite_fields}
  \end{figure}

Figure \ref{opposite_fields}(a, b, c) shows the evolution of the m$_z$ magnetization component in the nanowire throughout these steps, that ultimately lead to the creation of a DW state from an initial SV state. This method however has two drawbacks. The first one is the intensity of the applied magnetic field \diego{H$_0$}, which must be sufficient to be able to reverse the magnetization in one of the halves of NW while maintaining that of the other half. In this case, a limit value for the field amplitude of \diego{0.6 T} has been found, below which the final state remained the single vortex. The second drawback resides in the difficulty of creating strong magnetic fields with such a specific spatial asymmetry on a small scale. 

\section{BP-DW created by pinning}
\label{appendix:BP-DW_created_by_pinning}
In order to reduce the amplitude of the magnetic field necessary to create the BP-DW in the NW 
the introduction of BP-DW pinning has been investigated. This method dwells in the addition of a defect in the system trying to localize the BP-DW on it. In this case, a layer of non-magnetic material has been introduced in the middle plane of the nanowire \diego{(see modes of the system in Figure \ref{modes}d)}. Here, the same procedure has been followed as in the opposite magnetic fields method with the exception that a non-magnetic layer with M$_{s}$ = 0 and thickness of 2.5 nm has been introduced in the center of the 250 nm long nanowire. We found similar results as in the opposite magnetic fields method without pinning \diego{(see Appendix \ref{appendix:opp_fields})}, however, the value of the critical magnetic field needed for the DW to be generated was reduced from \diego{0.6 T} to \diego{0.45 T}.

Although the pinning of BP-DW we explored approaches us to the appropriate goal, it still presents drawbacks similar to the opposite magnetic fields method. One could try to reduce the critical field to generate the BP-DW even more by adding a thicker pinning layer, but high accuracy in the symmetry of the applied opposite fields would still be needed. Besides, as we describe in Appendix \ref{appendix:pinning}, extra pinning could complicate the goal of achieving low energy manipulation of the magnetic SV-BP topologies. In the next section we describe what we believe is a most practical way to realize the SV-BP transition, that is by injection of opposite spin currents.

\begin{figure}[h]
  \centering
  \includegraphics[width=1.0\linewidth]{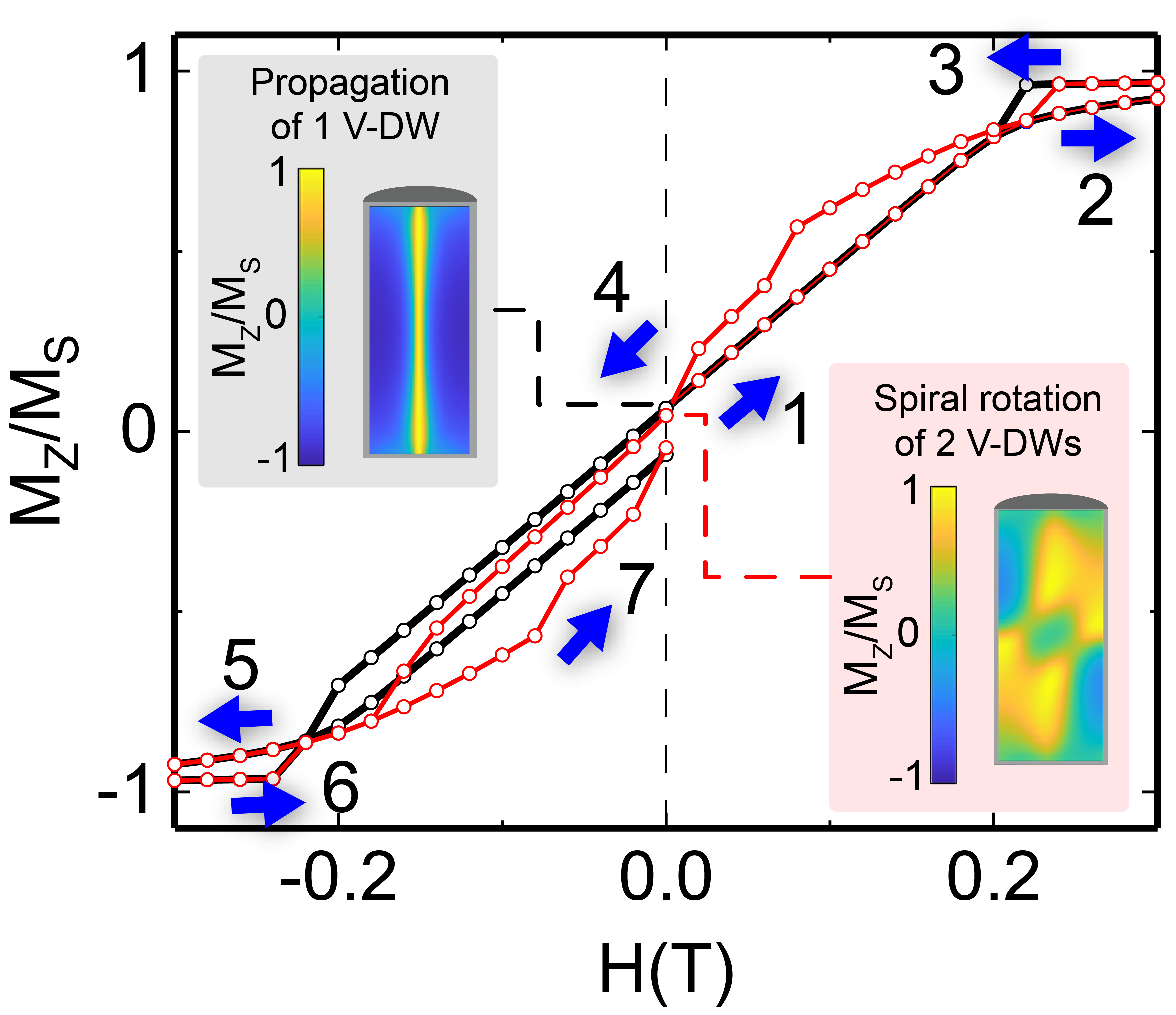}
  \caption{\diegotwo{Hysteresis loops for a 250 nm long NW in the case of each of the two presented reversal modes. Insets show the cross-section of the NW for two different types of detected magnetization reversal modes at H=0: spiral rotation of two V-DWs (2 V-DW) and propagation of one V-DW (1 V-DW).}} 
  \label{reversal_modes}
  \end{figure}

\begin{figure*}[t]
  \centering
  \includegraphics[width=1\linewidth]{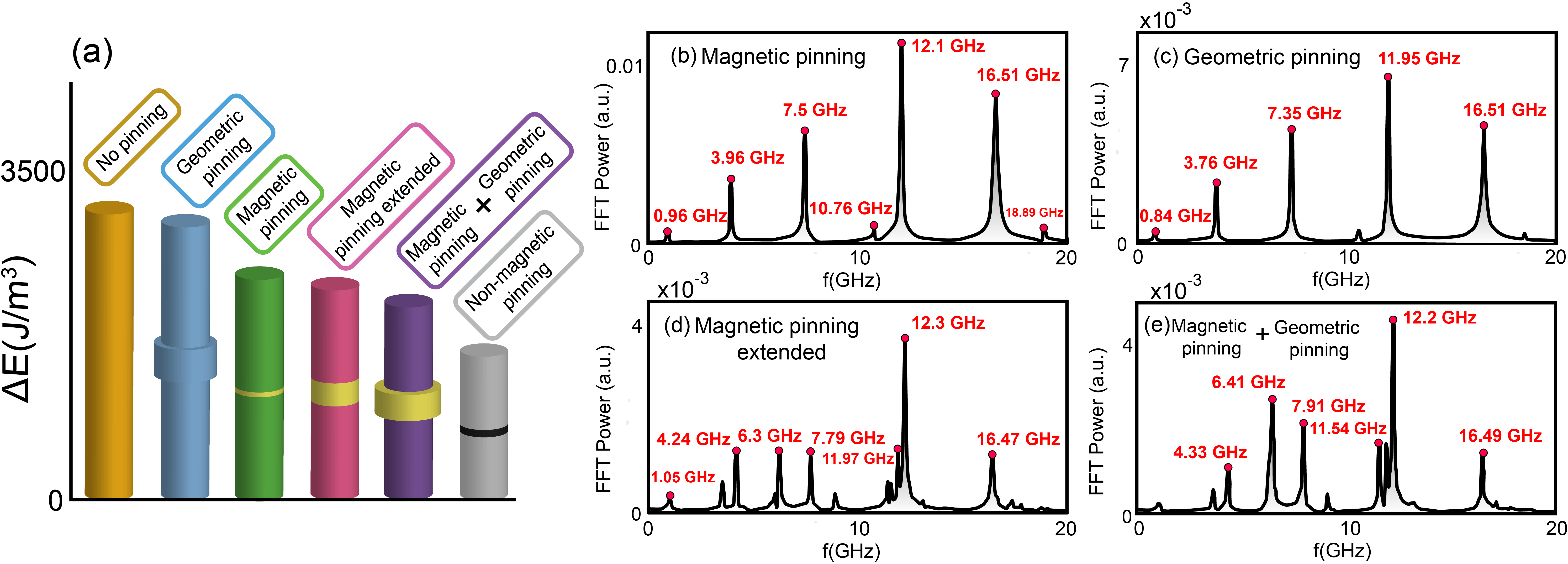} 
  \caption{(a) Energy gap between the BP-DW and the SV state for the different pinning introduced in the NW. Average spin wave modes spectra of the different pinned systems are depicted in (b, c, d, e).}
  \label{pinning}
  \end{figure*}

\section{Estimation of the reversal modes and their probability calculated from the obtained hysteresis loops}
\label{appendix:reversalmodes}

\diegotwo{To obtain the data presented in the inset of Figure \ref{tiempo_destruct}a in the main text, several hysteresis loops were performed saturating up to 1T of axial field (see Figure \ref{reversal_modes}). At H=0, when the field is being reduced, the static magnetic configuration reveals the reversal modes of the system. This process was executed for the NW lengths within the range of 220-280 nm depicted in the inset of Figure \ref{tiempo_destruct}a.}

 However, the type of reversal mode in this range was found to be uncertain, i.e., the NW hysteresis loop in axial magnetic field is characterized by a particular reversal mode (either the spiral rotation of two V-DWs or the propagation of one V-DW) each time a hysteresis loop was performed for a specific length. \diegotwo{Meaning that different reversal modes can be presented for the same system in separated hysteresis runs.} This is due to the fact that the two observed reversal modes are close in energy for most of the studied NW lengths. In order to describe the probabilistic character of the reversal mode topology, repeated simulations of ten hysteresis loops were performed for each NW length. The inset of Figure \ref{tiempo_destruct}a represents the probability to obtain the 1 V-DW propagation reversal mode vs the length of the NW. 

\section{Influence of pinning on the control over the magnetization topology}
\label{appendix:pinning}

The investigated BP-DW is formed as a metastable state above SV ground state (Fig. \ref{energies}). This facilitates the switching from the BP-DW to SV state under rather small fine tuned microwave field as well as the backward switching by using reasonable spin current densities (Fig.\ref{tiempo_destruct}). We have investigated the efficiency of such switching not only by varying the NW length (see Fig. \ref{tiempo_destruct}) but also through the implementation of different strategies for pinning of the BP-DW for the specific NW length of 250 nm for which switching of the magnetic texture topology was found to be less energetically costly. Specifically, we introduced BP-DW pinning in such NWs by using a modulated bamboo-like geometry \cite{Berganza2016,Saez2022} (geometric pinning), enhanced local saturation magnetization $M_s$ (magnetic pinning) or the implementation of a non-magnetic layer (non-magnetic pinning). We found that the energy gap between the two metastable states is reduced if the BP-DW position pinning is introduced (see Fig. \ref{pinning}a), as a consequence of an enhanced stability of the BP-DW state under the different pinning conditions. Our simulations reveal that the BP-DW stabilization is better achieved through a magnetic pinning, rather than geometrically modulating the middle of the NW, even though the geometrical modulation is introduced in 25 nm NW segment and the magnetic pinning is introduced only in a 2.5 nm segment.

Pinning of the BP position is an additional obstacle to perform the state switching via a microwave field. Thus, \diego{MW} excitation of the main SW modes in the studied amplitude range (up to \diego{2 mT}) could only destroy the BP-DW state in the case of the bamboo-like pinning, with the times of transition being only slightly above those of an ideal case 250 nm length NW (see Fig. \ref{tiempo_destruct}b). For any of other implemented pinning methods (see Fig. \ref{pinning}a), the transition from BP-DW to SV state is forbidden.

\section{\diegotwo{Latency times in the high MW excitation limit}}
\label{appendix:latency}

\begin{figure}[t]
  \centering
  \includegraphics[width=0.8\linewidth]{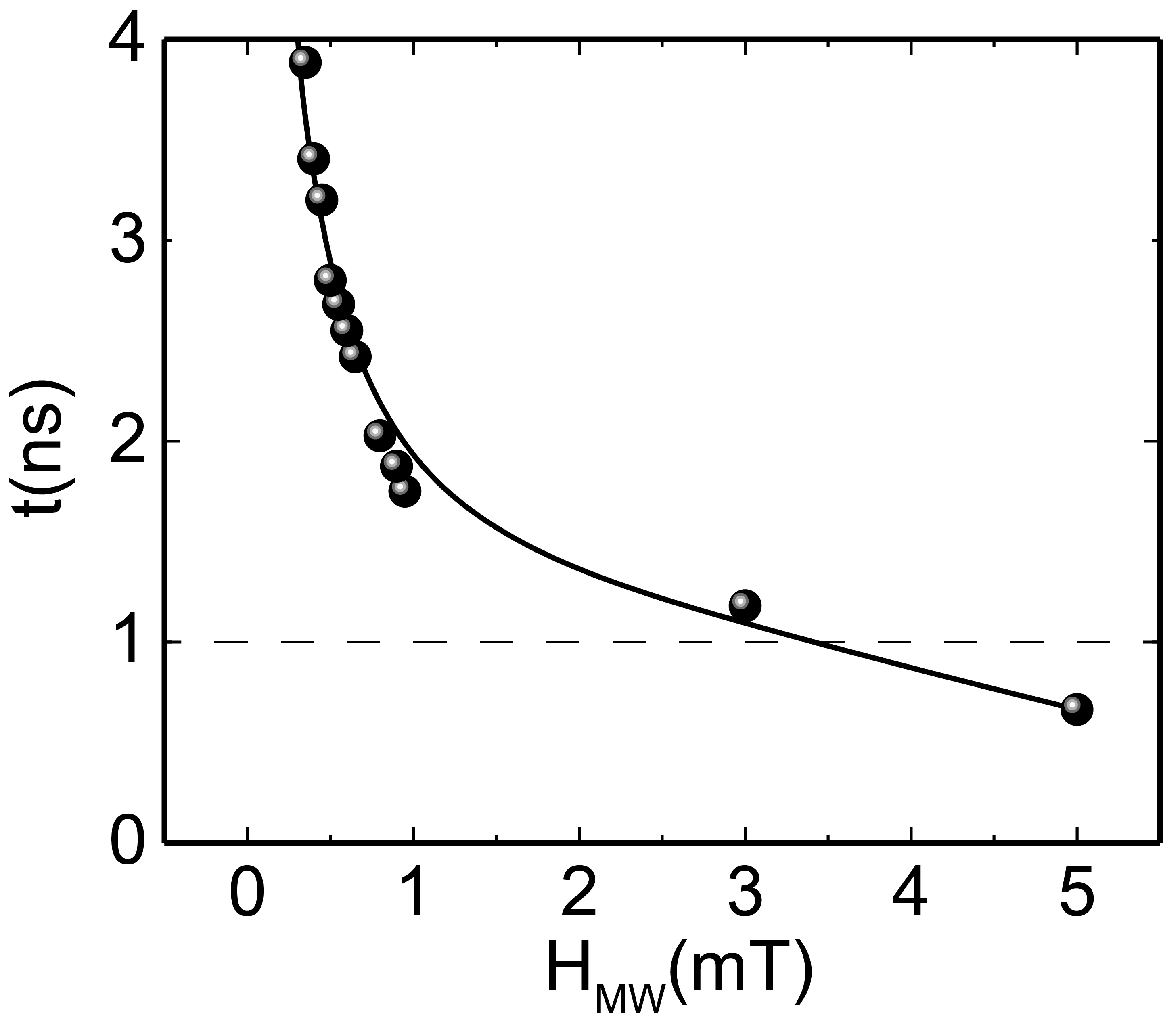} 
  \caption{\diegotwo{Time of destruction of the BP-DW state (latency time) in a 250 nm NW under a frequency-tuned to the main spin wave mode observed (11.9 GHz) MW excitation against the amplitude of the driving field.} }
  \label{latency_time}
  \end{figure}

\diegotwo{Latency times below the nanosecond are achieved in the BP-DW to SV state transition for high MW driving fields (see Figure \ref{latency_time}).}

\bibliography{bibliography}

\end{document}